\documentclass[12pt]{article}
\usepackage{epsfig,amsfonts,amssymb}
\usepackage{amsmath,graphics}

\usepackage{hyperref}
\def\ZZZ{{\hbox{ Z\kern-1.6mm Z}}}
\def\RRR{{\hbox{ R\kern-2.4mm R}}}
\def\CCC{{\hbox{ C\kern-2.0mm C}}}
\def\zzz{{\hbox{z\kern-1mm z}}}
\newcommand{\mathsym}[1]{{}}

\newcommand{\nn}{\nonumber \\}

\newcommand{\qeq}{{\hbox{=\kern-2.3mm ? \kern.5mm }}}
\renewcommand{\qeq}{=}

\newcommand{\be}{\begin{equation}}
\newcommand{\ee}{\end{equation}}
\newcommand{\ben}{\begin{eqnarray}\displaystyle}
\newcommand{\een}{\end{eqnarray}}

\newcommand{\sectiono}[1]{\section{#1}\setcounter{equation}{0}}

\def\one{{\hbox{ 1\kern-.8mm l}}}
\def\zero{{\hbox{ 0\kern-1.5mm 0}}}

\begin{document}

\baselineskip 24pt

\begin{center}
{\Large \bf On the smoothness of multi-M2 brane horizons
}

\end{center}

\vskip .6cm
\medskip

\vspace*{4.0ex}

\baselineskip=18pt

\centerline{\large \rm   Chethan N. Gowdigere, Siddharth Satpathy and
Yogesh K. Srivastava}

\vspace*{4.0ex}

\centerline{\large \it National Institute of Science Education and Research.}

\centerline{\large \it  Sachivalaya Marg, PO: Sainik School,}

\centerline{\large \it  Bhubaneswar 751005, INDIA}

\vspace*{1.0ex}
\centerline{E-mail: chethan.gowdigere,yogeshs@niser.ac.in,siddharthsatpathy.ss@gmail.com}

\vspace*{5.0ex}

\centerline{\bf Abstract} \bigskip

We calculate the degree of horizon smoothness of multi- $M2$-brane solution with branes along a common axis. We find that the metric is generically only thrice continuously differentiable at any of the horizons. The four-form field strength is found to be only twice continuously differentiable. We work with Gaussian null-like co-ordinates which are obtained  by solving geodesic equations for multi-$M2$ brane geometry. We also find different, exact co-ordinate transformations which take the metric from isotropic co-ordinates to co-ordinates in which metric is thrice differentiable at the horizon. Both methods give the same result that the multi-$M2$ brane metric is only thrice differentiable at the horizon.
\vfill \eject

\baselineskip=18pt

\tableofcontents

\sectiono{Introduction} \label{s1}

Multi-black hole solutions in four dimensional Einstein-Maxwell theory have been known and analyzed extensively \cite{Hartle} in literature.   In \cite{Candlish1}, smoothness of multi-black hole spacetimes in higher dimensional Einstein-Maxwell theory was analyzed using Gaussian null co-ordinates, building on earlier work by \cite{Welch}. In \cite{Candlish2}, similar analysis was done for multi-BMPV black holes. 
In this work we analyze the issue of horizon smoothness for the case of $M2$ brane of M-theory. Among the fundamental objects in string and M-theories, $M2$ branes \cite{Duff} are special in having an analytic horizon. In \cite{Gibbons}, multi-M2 brane metric was given and the possibility that horizon may not be analytic was mentioned. In \cite{Perry}, authors considered the case of infinite array of $M2$ branes and found that horizon is analytic in this case. This case is parallel to the higher dimensional black holes in compactified space-times, considered in \cite{Myers}. In this work we explore the problem of degree of smoothness of multi M2 brane horizon in detail. 
We use Gaussian null co-ordinates to construct a co-ordinate chart that covers the horizon. The metric near the horizon is constructed as a series in affine parameter along a null geodesic and the resulting metric coefficients are found to be thrice differentiable only. We also construct exact $\mathcal{C}^3$ extension of the metric by means of suitable co-ordinate transformations. Both approaches give the same result. Because of finite differentiability at horizon, the extension and hence the interior metric is not unique. 

\sectiono{Single centered $M2$ brane} \label{s2}

First we analyze the case of single centered M2 brane and find co-ordinates in which metric is analytic at the horizon and can be continued to the interior. Here the term analytic refers to real analyticity i.e being infinitely differentiable at the horizon as real functions and the existence of a Taylor series expansion around the horizon which converges to the function. As a matter of notation, we will call a function $f(x)$ as $\mathcal{C}^k$ function if function $f(x)$ and it's derivatives $f',f'', \ ... \ f^k$ exist and are continuous at the point in consideration.
The single or coincident $M2$ brane metric is 
\begin{eqnarray}
 ds^2 = H^{-2/3}(-dt^2 + dx_{1}^2 + dx_{2}^2 ) + H^{1/3}(dr^2 + r^2 d\Omega_{7}^
{2}) \\
C_{012} = H^{-1} \ , \ \ H = 1+ \frac{\mu_{1}}{r^6} 
\end{eqnarray}

To go to non-singular co-ordinates,we define, following \cite{Stelle}
\be
r = (\mu_{1})^{1/6}(\rho^{-3} -1)^{-1/6} = (\mu_{1})^{1/6}\sqrt{\rho}(1-\rho^3 )^{-1/6}
\ee

With this,we have $H=\frac{1}{\rho^3}$ and metric becomes
\begin{eqnarray}
\label{singlecenteredmetric}
ds^{2} = \rho^2 (-dt^2 + dx_{1}^2 + dx_{2}^2 )+(\mu_{1})^{1/3} (\frac{d\rho^2}{4\rho^2} + 
d\Omega_{7}^{2}) + \\ \nonumber
(\mu_{1})^{1/3}\frac{d\rho^2}{4\rho^2} \left[ (1- \rho^{3})^{-7/3} -1\right] + (\mu_{1})^{1/3} \left[
(1- \rho^{3})^{-1/3} -1\right]d\Omega_{7}^{2}
\end{eqnarray}

First line above is the metric for $AdS_4 \times S^{7}$ in Poincare 
co-ordinates(locally) while the second line is regular at $\rho \rightarrow 0$.  By doing the expansion in terms of $\rho$ around the horizon, we see that only  positive integer powers of $\rho$ occur in the metric coefficients and hence single centered case has analytic horizon. Later in this section we show that, inspite of appearance, the $AdS_4$ part metric is also  analytic and we can find co-ordinate transformation which express this.  In terms of $\rho$, the horizon (which was at 
$r=0$ in previous isotropic co-ordinates) is at $\rho=0$ while asymptotic 
infinity corresponds to $\rho \rightarrow 1^{-}$. Now we derive explicit regular co-ordinates for $AdS_4$ which will be useful later on also.
\begin{equation}
ds^{2}_{AdS_4} = \rho^2 (-dt^2 + dx_{1}^2 + dx_{2}^2 )+(\mu_{1})^{1/3} (\frac{d\rho^2}{4\rho^2})= \rho^2 (dudv + dx_{2}^2 )+(\mu_{1})^{1/3} (\frac{d\rho^2}{4\rho^2})
\end{equation}

where we have defined light-cone co-ordinates $u,v= x_1 \pm t$. 
 First thing to note is that the horizon is not just $\rho=0$ but also $t \rightarrow \infty$. So to 
go to the horizon we need the limits 
\be
\rho \rightarrow 0 \  \ , \ \ u,v \rightarrow \infty
\ee

Now we define new co-ordinates 
\be
\label{coordinatetransformation}
v= -\frac{1}{V} \ \ , \ \  u= U + \frac{1}{VW^2} + \frac{X_{2}^{2}}{V} \ \ ,  \ \ \rho= \frac{\mu_{1}^{1/6}V W}{2} \ \ , \ \ x_{2}= \frac{X_{2}}{V}
\ee

In these co-ordinates, horizon is at $V=0$ while $W$ is finite at the horizon. Using these, we get 
\begin{equation}
ds^2_{AdS_4} = \frac{(\mu_{1})^{1/3}}{4} \left( W^2 (dUdV + dX_{2}^{2}) + \frac{dW^2}{W^2} \right)
\end{equation}

One can see that in these co-ordinates the metric for $AdS_4$ is regular.Rest of the metric \ref{singlecenteredmetric} is easily seen to be analytic at the horizon in these co-ordinates.

\section{Differentiability of two centred M2 brane horizon using axial null geodesics} \label{sec3}

In this section, we follow \cite{Candlish1} to do a quick calculation to determine the differentiability of one particular component of the metric. This will prepare us for the use of Gaussian null co-ordinates for the multi-centred $M2$ brane metric which is done in next section. Here we use the argument given in \cite{Candlish1} which proves the following : If the metric admits a $C^k$ extension  through one of the horizons, with $k \geq 2$, and that this extension admits a Killing vector field $V$  then $V$ must be $C^k$. We will apply this result to the isometries of the solution corresponding to space-time translations along the brane world volume i.e. $V$ is any of $\frac{\partial}{\partial t}, \frac{\partial}{\partial x_1}, \frac{\partial}{\partial x_2}$. The norm of $V$ is  $\propto H^{-2/3}$. Since both the metric and $V$ are $C^k$, it follows that $H^{-2/3}$ must also be $C^k$ through the horizon. We can use this to determine  an upper bound on $k$ by 
 considering $H^{-2/3}$ along an axial null geodesic. 

Studying axial geodesics are best done in a co-ordinate system that uses cylindrical polar co-ordinates $z, \sigma,\Omega_6$  in the transverse space. The metric for the two centred M2 brane solutions in cylindrical polar co-ordinates  is 
\begin{equation}
ds^2=H^{-2/3}(-dt^2 + dx_{1}^2 + dx_{2}^2 ) + H^{1/3}(dz^2+d\sigma^2 +\sigma^2 d\Omega_6^2)
\end{equation}
with \begin{equation}
H=1+\frac{\mu_1}{(\sigma^2+z^2)^3} + \frac{\mu_2}{\left(\sigma^ 2 + (z-a)^2\right)^3}
\end{equation}
Consider a future directed null geodesic approaching the origin along the positive $z$-axis. This geodesic has a non-trivial dependence on the affine parameter for only four co-ordinates, $t(\lambda),x_{1}(\lambda),x_{2}(\lambda)$ and $z(\lambda)$; all other co-ordinates take constant values except $\sigma(\lambda) = 0$. Non-trivial geodesic equations are 
\be \label{1} \frac{d}{d\lambda} \left( -H^{-\frac23} \, \dot{t} \right)  = 0, \quad  \frac{d}{d\lambda} \left( H^{-\frac23} \dot{x_1} \right)  = 0, \quad \frac{d}{d\lambda} \left( H^{-\frac23} \dot{x_2} \right)  = 0 \ee

\be \label{2}  H^{-\frac23}\,(- \dot{t}^2 +  \dot{x_1}^2 + \dot{x_2}^2  ) + H^\frac13 (\dot{z}^2)  = 0 \ee

Here $\lambda$ derivatives are denoted by dot over the respective variables. In this special case, these geodesic equations can be integrated once\footnote{see next section for more details about this} to give 
\begin{eqnarray}
\label{eleclul}
 - H^{2/3} +H^{1/3}\dot{z}^2 = 0.
\end{eqnarray}
which result in  
\begin{equation}
\label{pikachu1}
\frac{dz}{d\lambda}=-H^{1/6} = -\left(1+\frac{\mu_1}{z(\lambda)^6} + \frac{\mu_2}{(z(\lambda)-a)^6}\right)^{\frac16}
\end{equation}
We know that the horizon is located at $z = 0$. We can choose that the affine parameter takes the value zero at the horizon so that $z(\lambda)$ is small near $\lambda = 0$. A small $z$ expansion of $H^{1/6}$ will have $1/z$ as the leading order term. Hence $\frac{dz}{d\lambda} \sim -\frac{1}{z} $ and this implies that $z(\lambda)$ can be expanded in terms of powers of $\sqrt{-\lambda}$, the minus sign because $\lambda$ is negative outside the horizon and becomes zero at the horizon. i.e.  $z \longrightarrow 0^+$ as $\lambda \longrightarrow 0^{-}$. Hence we make the ansatz 
\begin{equation}
z(\lambda) = \sum_n c_n\,(\sqrt{-\lambda})^n
\end{equation}
Now, we use ~\ref{pikachu1} to determine the coefficients $c_n$:
\begin{eqnarray}
z(\lambda)=\sqrt{2}\mu_1^{1/12}\sqrt{-\lambda}+\frac{\sqrt{2}\left( a^6 + \mu_2\right)}{6 a^6 \mu_1^{5/12}}\left( \sqrt{-\lambda}\right)^7+\frac{16\mu_2}{9 a^7 \mu_1^{1/3}}\lambda^4 \nonumber \\
+\frac{28\sqrt{2}\mu_2}{5a^8\mu_1^{1/4}}\left( \sqrt{-\lambda}\right)^9 - \frac{896\mu_2}{33 a^9 \mu_1^{1/6}}\lambda^5 +\frac{56\sqrt{2}\mu_2}{a^{10}\mu_1^{1/12}}\left( \sqrt{-\lambda}\right)^{11} \nonumber \\
 + \frac{2688\mu_2}{13a^{11}}\lambda^6 -\frac{17a^{12}+34a^6\mu_2 + 17\mu_2^2 -25344\,\mu_1 \mu_2}{36\sqrt{2}a^{12}\mu_1^{11/12}}  \left( \sqrt{-\lambda} \right)^{13} \nonumber \\
+ \frac{704\mu_2(a^6-216\mu_1 +\mu_2)}{135a^{13}\mu_1^{5/6}}\lambda^7 + \mathcal{O}\left( (-\lambda)^{15/2} \right)
\end{eqnarray}
The norm of the Killing vector fields is then computed: 
\begin{equation}
\label{bokwa}
H^{-2/3}=\frac{4\lambda^ 2}{\mu_1^{1/3}} + \frac{56 (a^6 + \mu_2)}{3 a^6 \mu_1^{5/6}} \lambda^5 - \frac{1024\sqrt{2}\mu_2}{9a^7\mu_1^{3/4}}(\sqrt{-\lambda})^{11}+\mathcal{O}\left( \lambda ^6 \right)
\end{equation}

We thus see that the norm is a  $\mathcal{C}^5$ and  not $\mathcal{C}^6$ function;  the addition of an extra centre has decreased the horizon smoothness.  Also, the single centered case can be obtained by substituting $\mu_2=0$ in the above and we see that the metric is analytic (at least to the order we have have displayed) but of course we know this from the  construction of exact co-ordinates in the previous section. 

One can try to consider norm of Killing vectors corresponding to the $SO(7)$ symmetry of the metric to determine the degree of differentiability of corresponding metric components. But the six sphere  along  which these Killing vector fields are supported become zero size everywhere on the axial geodesic and consequently the norms of the Killing vector fields vanish. We would need to consider radial geodesics which we do in the next section.

\section{Gaussian null-like  co-ordinates}

In this section we will construct a co-ordinate system which will provide a good co-ordinate system in the neighbourhood of the event horizon. The co-ordinate system is obtained from the family of non-axial null geodesics. In this sense it is similar to the Gaussian null co-ordinate system for the neighbourhood of an event horizon. But the co-ordinate system we construct is not exactly the Gaussian null co-ordinate system; on the horizon hypersurface one of the Guassian null co-ordinates is the affine parameter along the null geodesic generators of the horizon but the co-ordinates we construct below do not have this feature. Nevertheless it does provide a good co-ordinate system for the neighbourhood of the horizon which is what is needed to address the smoothness of the metric at the horizon and to extend it into the interior.  We will refer to the co-ordinate system we construct as Gaussian null-like co-ordinates.  
  
The family of non-axial null geodesics are best studied in a spherical co-ordinate system for the transverse space to the branes. The multi-centered $M2$ brane solution in these co-ordinates is given by

\be \label{im} ds^2  =  H^{-\frac23}\,(-dt^2 + dx_1^2 + dx_2^2) + H^\frac13\,(dr^2 + r^2 d\theta^2 + r^2\sin^2 \theta\,d\Omega_6^2 ) \ee
with 
\be H = 1+ \frac{\mu_1}{r^6} + \sum_{i = 2}^{N}\frac{\mu_i}{(r^2 - 2 a_i r \cos \theta + a_{i}^2)^3}.\ee
The horizon of the first centre is at $r = 0$ and we will investigate the smoothness there. It is convenient to expand in terms of the relevant spherical harmonics
\begin{equation}
H=\frac{\mu_1}{r^6}+\sum_{n=0}^\infty h_n r^n\, Y_n(\cos \theta)
\end{equation}
where the $Y_n(\cos \theta)$ are certain Gegenbauer polynomials
\begin{eqnarray}
Y_0(\cos \theta)&=&1 \nonumber \\
Y_1(\cos \theta)&=& 6 \cos \theta \nonumber \\
Y_2(\cos \theta)&=& 24 \,\cos ^2\theta -3 \nonumber \\
Y_3(\cos \theta)&=& 80 \, \cos ^3 \theta  - 24 \cos \theta  \nonumber \\
Y_4(\cos \theta)&=& 240 \, \cos ^4 \theta  -120 \, \cos ^2 \theta  + 6\nonumber \\
Y_5(\cos \theta)&=& 672 \, \cos ^5 \theta - 480 \, \cos ^3 \theta  + 60 \,\cos \theta \nonumber \\
Y_6(\cos \theta)&=& 1792 \,\cos ^6 \theta -1680 \, \cos ^4 \theta  + 360 \,\cos ^2 \theta  -10 \nonumber 
\end{eqnarray}
and the coefficients $h_n$ are 
\be h_n = \delta_{n,0} + \sum_{i=2}^{N}\frac{\mu_i}{a_{i}^{6+n}}.\ee

\subsection{Constructing Gaussian null-like co-ordinates}

We expect that the $SO(7)$ symmetry of the solution outside the horizon continues to be a symmetry of the extension of the metric.  Hence we will only need to consider geodesics that have  constant angular momentum along the $S^6$ i.e. angular co-ordinates along $S^6$ will not change along the geodesic. The co-ordinates that are non-trivial functions of the affine parameter are $t, x_1, x_2, r, \theta$. The geodesic equations in isotropic co-ordinates then are five coupled second-order differential equations four of which can be integrated once :
\be \label{1} \frac{d}{d\lambda} \left( -H^{-\frac23} \, \dot{t} \right)  = 0, \quad  \frac{d}{d\lambda} \left( H^{-\frac23} \dot{x_1} \right)  = 0, \quad \frac{d}{d\lambda} \left( H^{-\frac23} \dot{x_2} \right)  = 0 \ee

\be \label{2}  H^{-\frac23}\,(- \dot{t}^2 +  \dot{x_1}^2 + \dot{x_2}^2  ) + H^\frac13 \left( \dot{r}^2  + r^2\, \dot{\theta}^2\right) = 0 \ee

\be \label{3}\ddot{r}  - \frac{\partial_rH}{3\,H}  + \frac{\partial_r H}{6 H} \left( \dot{r}^2 - r^2\, \dot{\theta}^2\right) - r \,\dot{\theta}^2 + \frac{\partial_\theta\,H}{3\,H}\,\dot{r}\,\dot{\theta} = 0  \ee

The angles on the $S^6$ do not change along the geodesic; they continue to be co-ordinates in the   Gaussian null-like co-ordinate  system. The angle $\theta$ changes along the geodesic; it's value on the horizon, denote it by  $\Theta$,  is taken to be one of Gaussian null-like co-ordinates. The other four co-ordinates in the Gaussian null-like co-ordinate system are the affine parameter $\lambda$ and the integration constants that come by integrating the $t, x_1, x_2$ geodesic equations.
We first solve the equations \eqref{1}  in the following manner :
\be \label{4.30} t = v - f(v,w,y_2)T(\lambda, \Theta),\quad  x_1 = w - g(v,w,y_2)\, T(\lambda, \Theta),  \quad x_2 = y_2 - h(v,w,y_2)\, T(\lambda, \Theta) \ee
where \be \label{T} T(\lambda, \Theta) \equiv \int d\lambda \,H^\frac23(\lambda, \Theta) \ee  will be determined below; the integration constants $v, w, y_2$ will form three of the  Gaussian null-like co-ordinates. Isotropic co-ordinates $t,x_1,x_2$ go bad as we approach horizon. This can be seen from \ref{4.30} where $T(\lambda, \Theta) \rightarrow \infty$ as $\lambda \rightarrow 0$. Gaussian null-like co-ordinates $v,w,y_2$ are well-defined and 
 finite at the horizon. 

 We choose to introduce the functions $f, g, h$ of integration constants  in the above  because a simpler choice such as constants will not make the metric non-singular at the horizon; then we would not have obtained a good co-ordinate system for the neighbourhood of the  horizon. A completely arbitrary choice of functions $f, g, h$ does not make the metric in these co-ordinates non-singular. We will encounter various conditions along the way one of them is that they need to satisfy  the constraint
\be \label{fgh} S \equiv f^2 - g^2 - h^2 -1 =0 \ee 
Although we do not have a solution to all the constraints\footnote{There are two of them only : \eqref{fgh} and \eqref{nonsing}.}  that the $f, g, h$ would need to satisfy by the end of the analysis, we do have many examples; thus we do have multiple examples of Gaussian null-like co-ordinate systems. The constraint \eqref{fgh} means that we now need to solve the equations 
\be \label{4}  - H^{\frac23} + H^\frac13 \left( \dot{r}^2  + r^2\, \dot{\theta}^2\right) = 0 \ee

\be \label{5}\ddot{r}  - \frac{\partial_rH}{3\,H}  + \frac{\partial_r H}{6 H} \left( \dot{r}^2 - r^2\, \dot{\theta}^2\right) - r \,\dot{\theta}^2 + \frac{\partial_\theta\,H}{3\,H}\,\dot{r}\,\dot{\theta} = 0.  \ee
Let us assign the affine parameter $\lambda = 0$ for the event horizon; here $\lambda$ takes positive values outside the horizon (note that this convention is different from the previous section). $r( \lambda, \Theta)$ and $\theta( \lambda, \Theta)$ are solutions to \eqref{4} and \eqref{5} with the initial conditions 
\be \label{bc}  r(0, \Theta) = 0, \quad \theta(0, \Theta) = \Theta, \quad \dot{\theta}(0, \Theta) = 0. \ee
We assume series expansions of the form
\be \label{series} r(\lambda, \Theta) = \sum_{n = 1} c_n(\Theta) \, \lambda^{\frac{n}{2}}, \qquad \theta(\lambda, \Theta) = \Theta + \sum_{n = 1} b_n(\Theta) \, \lambda^{\frac{n}{2}},\ee
the $\lambda^\frac12$ expansion parameter again coming from considering the leading order behaviour of the geodesic equations at $\lambda = 0$. The result of solving \eqref{4} and \eqref{5} order by order is  
\be \label{9} r(\lambda, \Theta) = \sqrt{2}\mu_ 1^{1/12}\,\lambda^{1/2}+\frac{h_0}{3\sqrt{2}\mu_1^{5/12}}\lambda^{7/2}+\frac{16\,h_1}{9\mu_1^{1/3}}\, c_\Theta \,\lambda^ 4 + \frac{4\sqrt{2}\,h_2}{5\,\mu_1^{1/4}}\,(3 + 4 c_{2 \Theta})\,\lambda^{9/2} + \mathcal{O}(\lambda^5) \ee
\be \label{10} \theta(\lambda, \Theta) = \Theta -\frac{24\sqrt{2}h_1}{35\mu_1^{5/12}} s_\Theta \,\lambda^{7/2}- \frac{4 h_2\,}{\mu_1^{1/3}} s_{2 \Theta}\,\lambda^4 -\frac{32\,\sqrt{2}\,h_3}{21\mu_1^{1/4}}\,(3 s_\Theta +5 s_{3\Theta})\,\lambda^{9/2} + \mathcal{O}(\lambda^5) \ee
where $c_\alpha\equiv\cos \alpha, s_\alpha \equiv \sin \alpha$. The appendix provides more details of obtaining these expansions as well as  more  terms all of which are needed to obtain the results in this section. With these expansions the computation in \eqref{T} can be performed. We thus have, in 
\eqref{4.30},  \eqref{9} and \eqref{10} the co-ordinate transformations between the Gaussian null-like co-ordinates $v, w, y_2, \lambda, \Theta, \Omega_6$ and the isotropic co-ordinates $t, x_1, x_2,  r, \theta, \Omega_6$. The functions $f, g, h$ thus enter in the definition of the Gaussian null-like co-ordinates; they only need to satisfy the constraint \eqref{fgh} and any other conditions that ensures the metric on the horizon in the Gaussian null-like co-ordinates is non-singular. We will see below that we have many choices for $f, g, h$.

\subsection{Metric is $\mathcal{C}^3$}
We use the co-ordinate transformations \eqref{4.30}, \eqref{9} and \eqref{10} to compute the metric in the Gaussian null-like co-ordinates : 
\begin{eqnarray} \label{c3metric} ds^2 = H^{-\frac{2}{3}}(\lambda,\Theta)\Bigg[\Big[-1+2T(\lambda ,\Theta) \, \partial_v f + T^2(\lambda, \Theta) \, z_1(v,w,y_2) \Big] \, dv^2 \nonumber \\
+\Big[1-2T(\lambda ,\Theta) \, \partial_w g + T^2(\lambda, \Theta) \, z_2(v,w,y_2) \Big] \, dw^2 + \Big[1-2T(\lambda ,\Theta) \, \partial_{y_2} h + T^2(\lambda, \Theta) \, z_3(v,w,y_2) \Big] \, dy_2^2 \Bigg] \nonumber \\
+2H^{-\frac{2}{3}}(\lambda,\Theta) \, T(\lambda,\Theta)\Bigg[\Big[q_1(v,w,y_2)+T(\lambda,\Theta)q_2(v,w,y_2)\Big] \, dv \, dw \nonumber \\ 
+ \Big[q_3(v,w,y_2)+T(\lambda,\Theta)q_4(v,w,y_2)\Big] \, dv \, dy_2 + \Big[q_5(v,w,y_2)+T(\lambda,\Theta)q_6(v,w,y_2)\Big] \, dw \, dy_2 \Bigg] \nonumber \\
 +2f \, dv \, d\lambda -2g \, dw \, d\lambda -2h \, dy_2 \, d\lambda +2H^{-\frac{2}{3}}(\lambda,\Theta)\partial_{\Theta}T(\lambda, \Theta) \Bigg[f\, dv \, d\Theta -g\, dw \, d\Theta -h \, dy_2 \, d\Theta \Bigg]\nonumber \\ 
+ \mathcal{O}(\lambda^8)\,d\lambda^2+ H^{\frac{1}{3}}(\lambda,\Theta)r^2(\lambda, \Theta) \, d\Theta^2  + \mathcal{O}(\lambda^7) \, d\lambda\,d\Theta +  H^{\frac{1}{3}}(\lambda,\Theta)r^2(\lambda, \Theta)\sin ^2(\theta(\lambda,\Theta))   \,d\Omega_6^2 \ \ \ \ \ \  \end{eqnarray}
Each of the unknown functions appearing above are defined in an appendix: \eqref{b1}, \eqref{b2}.  The condition $S=0$ \ref{fgh} ensures that $g_{\lambda\lambda}$ component is well-behaved at the horizon. Derived conditions $\partial_v S=0,\partial_w S=0, \partial_{y_2} S=0 $ ensure that 
$g_{v\lambda},g_{w\lambda},g_{y_2 \lambda}$ respectively are well-behaved at the horizon.

As we can see from the explicit components of the metric, the least differentiable components of the metric are $g_{\Theta \Theta}$ and $g_{\Omega_6 \Omega_6}$ which are  $\mathcal{C}^3$ functions. Hence we conclude that multi-centred membrane solution is only $\mathcal{C}^3$ at any of it's horizons. 

The metric \eqref{c3metric} has the  usual co-ordinate singularities of spherical co-ordinates at $\Theta=0,\pi$. Apart from these, we also require that the metric is non-singular at $\lambda = 0$. We can compute the determinant of the metric on the horizon i.e. at $\lambda = 0$:
\ben \label{det} g = \frac{\mu_1^{\frac43}}{16} \sin^{12} \Theta\, g_{S^6}\, [ f^2 (q_6^2 - z_2 z_3) + g^2 (q_4^2 - z_3 z_1) + h^2 (q_2^2 - z_1 z_2)   \nonumber \\ + 2 f g (q_4 q_6 - q_2 z_3) - 2 g h (q_2 q_4 - q_6 z_1) +  2 f h (q_6 q_2 - q_4 z_2)   ]\een
where $g_{S_6}$ is the determinant of the round metric on the unit six sphere and  the $q_i$'s and the $z_i$'s are defined in \eqref{b2}. Requiring that the determinant does not vanish on the horizon gives us the following condition that our choice of $f, g, h$ functions must satisfy:
\be \label{nonsing}   f^2 (q_6^2 - z_2 z_3) + g^2 (q_4^2 - z_3 z_1) + h^2 (q_2^2 - z_1 z_2)   + 2 f g (q_4 q_6 - q_2 z_3) - 2 g h (q_2 q_4 - q_6 z_1) +  2 f h (q_6 q_2 - q_4 z_2)  \neq 0. \ee. Simple algebraic manipulations give that this is equivalent to 
\be
(f\sqrt{(-q_6^2 + z_2 z_3)} - g\sqrt{(-q_4^2 + z_1 z_3)} -h\sqrt{(-q_2^2 + z_2 z_1)})^2  \neq 0
\ee

We do not analyze this condition in detail here. But we do have many examples for the $f, g, h$ functions that satisfy \eqref{nonsing} and \eqref{fgh} two of which are
\begin{eqnarray}
f(v, w,y_2) &=&\frac{1}{2}\left( w + \frac{1}{w} + \frac{y_2^2}{w} \right), g(v, w,y_2)= \frac{1}{2}\left( -w + \frac{1}{w} + \frac{y_2^2}{w} \right), h(v, w,y_2)= y_2. \nonumber \\
f(v, w,y_2)&=&\sqrt{1+y_{2}^{2}}\cosh w ,\quad g(v, w,y_2)= \sqrt{1+y_{2}^{2}}\sinh w , \quad h(v, w,y_2)= y_2.
\end{eqnarray}

\subsection{Volume of $S^6$}
In the previous section we deduced that the metric is only $\mathcal{C}^5$ from the differentiability of the norm of the Killing vector field corresponding to time translations. This was achieved by examining the norm of the Killing vector field along the axial geodesic. This strategy does not work for computing the norm of the $SO(7)$ Killing vector fields because the six sphere on  which these Killing vector fields are supported becomes zero size everywhere on the axial geodesic and consequently the norms of the Killing vector fields vanish. By considering radial geodesics, we have obtained the metric in the neighbourhood of the event horizon. Using this metric, one can determine volume of the $S^6$:
\be A_6 = \mu_1\,\sin^6\Theta + 8\,h_0\,\mu_1^{\frac12}\, \sin^6 \Theta\,\lambda^3+ \frac{1536 \sqrt{2}}{35} h_1 \mu _1^{7/12} \, \cos \Theta  \sin ^6\Theta  \lambda ^{7/2}  + \mathcal{O}(\lambda^4).\ee 
We thus see that the volume is a $\mathcal{C}^3$ function. From the relation of the norm of the $SO(7)$ Killing fields  to the volume of the $S^6$ we can conclude that the norm is also only $\mathcal{C}^3$.

\subsection{Maxwell field strength is $\mathcal{C}^2$}

We now consider the degree of differentiability of $4$-form field strength in the Gaussian null-like co-ordinates.  The 3-form  gauge potential in the Gaussian-null co-ordinates is obtained using the co-ordinate transformations \eqref{4.30}, \eqref{9}, \eqref{10}: 
\begin{eqnarray}\label{mpot}
A_{[3]} &=& H^{-1}dt \wedge dx_1 \wedge dx_2 \nonumber \\
&=& H^{-1} \left( dv-fdT-Tdf \right) \wedge \left( dw-gdT -Tdg \right) \wedge \left( dy_2 -hdT -Tdh \right) 
\end{eqnarray}
The full expression can be found in an appendix: \eqref{c1}, \eqref{c21} and \eqref{c22}. The 4-form field strength in the Gaussian null-like co-ordinates is: 
\begin{eqnarray}
F_{[4]} &=& d\left(H^{-1}\right) \wedge dt \wedge dx_1 \wedge dx_2 \nonumber \\
&=& \left( \partial_{\lambda}H^{-1} \, d\lambda + \partial_{\Theta}H^{-1} \, d\Theta \right) \wedge \left( dv-fdT-Tdf \right) \wedge \left( dw-gdT -Tdg \right) \wedge \left( dy_2 -hdT -Tdh \right) \nonumber \\
\end{eqnarray}
The  non-zero components are the following:
\begin{eqnarray}\label{mfield}
F_{\lambda v w y_2} &=& \partial_{\lambda}H^{-1} \Big[ 1 - u_1(v,w,y_2) \, T  +  u_2(v,w,y_2) \, T^2 \Big] \nonumber \\
F_{\Theta v w y_2} &=& \partial_{\Theta}H^{-1} \Big[ 1 - u_1(v,w,y_2) \, T  +  u_2(v,w,y_2) \, T^2 \Big] \nonumber \\
F_{\lambda \Theta v w} &=& \left( \partial_{\lambda}H^{-1} \, \partial_{\Theta}T - \partial_{\lambda}T \, \partial_{\Theta}H^{-1} \right) \Big[ h(v,w,y_2) - u_3(v,w,y_2) \, T +  u_4(v,w,y_2) \, T^2 \Big] \nonumber \\
F_{\lambda \Theta v y_2} &=& \left( \partial_{\lambda}H^{-1} \, \partial_{\Theta}T - \partial_{\lambda}T \, \partial_{\Theta}H^{-1} \right) \Big[ g(v,w,y_2) - u_5(v,w,y_2) \, T +  u_6(v,w,y_2) \, T^2 \Big] \nonumber \\
F_{\lambda \Theta w y_2} &=& \left( \partial_{\lambda}H^{-1} \, \partial_{\Theta}T - \partial_{\lambda}T \, \partial_{\Theta}H^{-1} \right) \Big[ f(v,w,y_2) - u_7(v,w,y_2) \, T +  u_8(v,w,y_2) \, T^2 \Big] \nonumber \\
\end{eqnarray}
The various undefined terms above  are gathered in an appendix: \eqref{c21}, \eqref{c22}, \eqref{c3}. Conditions $\partial_v S=0,\partial_w S=0, \partial_{y_2} S=0 $ require
\begin{equation}
\left| \begin{array}{ccc}
\partial_v f & -\partial_v g & -\partial_v h \\
\partial_w f & -\partial_w g & -\partial_w h \\
 \partial_{y_2} f & -\partial_{y_2} g & -\partial_{y_2} h \\\end{array} \right| =0
\end{equation}
for having non-trivial solution for $f,g,h$. This condition ensures that components $F_{\lambda v w y_2},F_{\Theta v w y_2}$ are regular at the horizon.

>From the formulae in the appendix, particularly \eqref{c21} and \eqref{c22} , we can see that the $F_{\lambda v w y_2} $ component is a $\mathcal{C}^3$ function,  $F_{\Theta v w y_2}$ is a $\mathcal{C}^4$ function and $F_{\lambda \Theta v w}, F_{\lambda \Theta v y_2}, F_{\lambda \Theta w y_2}$ are $\mathcal{C}^2$ functions. We thus conclude that $F_{[4]}$ is only $\mathcal{C}^2$ at the horizon.

\subsection{Equation of motion}
\begin{equation}\label{eom}
R_{\mu \nu}=\frac{1}{12}\left(F_{\mu \alpha \beta \gamma}F_{\nu}^{\, \, \alpha \beta \gamma}-\frac{1}{12}F^2g_{\mu \nu} \right)
\end{equation}
We have shown that the metric is  $\mathcal{C}^3$. Roughly one then expects that the curvature components are $\mathcal{C}^1$ functions i.e. the left hand side of the equation of motion \eqref{eom} is $\mathcal{C}^1$. On the other hand we have shown that the Maxwell field strength is $\mathcal{C}^2$  which means roughly that the right hand side of \eqref{eom} have $\mathcal{C}^2$ components. We are thus left with a puzzle about the mismatch of differentiability of the right and left hand sides of the equation of motion. To settle this, we simply perform the computations of the left and right hand sides of \eqref{eom}. We choose a specific choice of Gaussian null-like co-ordinates \eqref{d1} and the results of the computations are collected in an appendix : \eqref{ricci}, \eqref{d5}.

>From an examination of the Ricci tensor components \eqref{ricci} it is clear that there are no components which are only $\mathcal{C}^1$ thus solving our puzzle. The components of the right hand side of \eqref{eom} have also been computed and they match with the Riccci tensor components as they should. 

\subsection{Extending through the horizon and the  interior metric}

Since the metric at the horizon is only of finite differentiability, infinitely many interior solutions can be matched to exterior solution given by \eqref{im}. We will assume that interior metric has same Killing vectors as exterior metric.   As done for the black hole case in \cite{Candlish1}, we assume that the interior metric takes the same form as exterior metric in  isotropic co-ordinates but with a different harmonic function $\hat{H}(r,\theta)$. The range of the co-ordinates (except $r$) remains the same as the exterior metric. 
\begin{equation}
ds^2=\hat{H}^{-2/3}(-dt^2+dx_1^2+dx_2^2)+\hat{H}^{1/3}(dr^2+r^2d\theta^2 +r^2sin^2\theta d\Omega_2^2)
\end{equation}

where $\hat{H}$ is chosen to agree with $H$ (for exterior metric) at the leading order. 
\begin{equation}
\hat{H}=\frac{\mu_1}{r^6}+\sum_{n=0}^\infty \hat{h}_n r^n Y_n(cos\theta)
\end{equation}

In the exterior region, we have $\lambda >0$ (this is consistent with section $4$ but not with section $3$) which is the affine parameter along a past-directed geodesic. For the interior region, we define the parameter $\hat{\lambda}$ to be the affine parameter along future directed null geodesic.   We construct (nearly) Gaussian null-like co-ordinates as before. Geodesic equation for $r$ and $\theta$ remain same except for $\lambda \rightarrow \hat{\lambda}$ and $h_{n} \rightarrow \hat{h}_{n}$.

\begin{equation}
t=v +f(v,w,y_2) \, \hat{T}\left(\hat{\lambda},\Theta \right), \ \ \ \ \hat{T}\left(\hat{\lambda},\Theta \right)\equiv \int \hat{H}(\hat{\lambda},\Theta)^{2/3}d\hat{\lambda}
\end{equation}

and similarly for $x_{1},x_{2}$ with $\hat{T}\left(\hat{\lambda},\Theta \right)$ defined as above. 
\be  x_1 = w +g(v,w,y_2) \, \hat{T}\left(\hat{\lambda},\Theta \right),  \quad x_2 = y_2 +h(v,w,y_2) \, \hat{T}\left(\hat{\lambda},\Theta \right) \ee
 The interior metric we get is
\begin{eqnarray} ds^2 = \hat{H}^{-\frac{2}{3}}(\hat{\lambda},\Theta)\Bigg[\Big[-1-2\hat{T}(\hat{\lambda} ,\Theta) \, \partial_v f + \hat{T}^2(\hat{\lambda}, \Theta) \, z_1(v,w,y_2) \Big] \, dv^2 \nonumber \\
+\Big[1+2\hat{T}(\hat{\lambda} ,\Theta) \, \partial_w g + \hat{T}^2(\hat{\lambda}, \Theta) \, z_2(v,w,y_2) \Big] \, dw^2 + \Big[1+2\hat{T}(\hat{\lambda} ,\Theta) \, \partial_{y_2} h + \hat{T}^2(\hat{\lambda}, \Theta) \, z_3(v,w,y_2) \Big] \, dy_2^2 \Bigg] \nonumber \\
+2\hat{H}^{-\frac{2}{3}}(\hat{\lambda},\Theta) \, \hat{T}(\hat{\lambda},\Theta)\Bigg[\Big[-q_1(v,w,y_2)+\hat{T}(\hat{\lambda},\Theta)q_2(v,w,y_2)\Big] \, dv \, dw \nonumber \\ 
+ \Big[-q_3(v,w,y_2)+\hat{T}(\hat{\lambda},\Theta)q_4(v,w,y_2)\Big] \, dv \, dy_2 + \Big[-q_5(v,w,y_2)+\hat{T}(\hat{\lambda},\Theta)q_6(v,w,y_2)\Big] \, dw \, dy_2 \Bigg] \nonumber \\
 -2f \, dv \, d\hat{\lambda} +2g \, dw \, d\hat{\lambda} +2h \, dy_2 \, d\hat{\lambda} -2\hat{H}^{-\frac{2}{3}}(\hat{\lambda},\Theta)\partial_{\Theta}\hat{T}(\hat{\lambda}, \Theta) \Bigg[f\, dv \, d\Theta -g\, dw \, d\Theta -h \, dy_2 \, d\Theta \Bigg]\nonumber \\ 
+ \mathcal{O}(\hat{\lambda}^8)\,d\hat{\lambda}^2+ \hat{H}^{\frac{1}{3}}(\hat{\lambda},\Theta)r^2(\hat{\lambda}) \, d\Theta^2  + \mathcal{O}(\hat{\lambda}^7) \, d\hat{\lambda}\,d\Theta +  \hat{H}^{\frac{1}{3}}(\hat{\lambda},\Theta)r^2(\hat{\lambda})\sin ^2(\theta(\hat{\lambda},\Theta))   \,d\Omega_6^2. \ \ \ \ \ \  \end{eqnarray}

For the metric to be continuous across the horizon, i.e. $\mathcal{C}^{0}$, we have to choose $\hat{\lambda}= -\lambda$ for the interior region.  We match interior and exterior metrics upto order $\lambda^3$ since we only have a $\mathcal{C}^3$ metric. This sets $h_{0}= -\hat{h}_{0}$ and rest of the coefficients are unconstrained. Hence we have infinite family of interior metrics parametrized by $h_{n}$ for $n \geq 1$.  Similarly field strength can be matched upto order $\lambda^2$ terms without any further constraints on $h_{n}$.

\section{Two centered M2-brane solution: Exact $C^{3}$ metric}

Metric for two centered M2 branes written in nearly Gaussian null co-ordinates provides a $C^3$ extension of the metric, written as a series in affine parameter $\lambda$. In this section, we give an exact $C^3$ metric for the two centered $M2$ brane case. Final form of the metric is not very illuminating though.

As we wrote earlier, for two M2 branes, one at the origin and other one along the $z$-axis at a distance $a$, we have

\begin{equation}
H = 1+\frac{\mu_{1}}{r^6} + \frac{\mu_{2}}{\left(r^ 2 + a^2 -2a r \cos\theta\right)^3}
\end{equation}

In section two, we had given co-ordinate transformations which make the metric of single centered $M2$ brane analytic at the horizon. For two $M2$-brane case, we won't get analytic metric but only $C^3$ metric. We start by making the same co-ordinate transformation as we use for single centered case. $r = \mu_{1}^{1/6}(\rho^{-3} -1)^{-1/6}$. Using this we get
$H= \frac{f(\rho,\theta)}{\rho^{3}}$ and metric becomes

\begin{equation}
ds^2 = \frac{\rho^2}{f^{2/3}} (dudv + dy_{2}^{2}) + f^{1/3}\mu_{1}^{1/3} \left( (1-\rho^3)^{-7/3}\frac{d\rho^2}{4\rho^2}  + (1-\rho^{3})^{-1/3}d\Omega_{7}^{2} \right)
\end{equation}

Here 
\begin{equation}
f(\rho,\theta) = 1+ \rho^3 \sum_{n=0}^{\infty} h_{n}Y_{n}(\cos\theta)(1-\rho^{3})^{-n/6}\rho^{n/2}
\end{equation}

and  $Y_{n}(\cos\theta)$ are Gegenbauer polynomials and $h_n$ are constants given by

\begin{equation}
h_n = \frac{\mu_{2}\mu_{1}^{n/6}}{a^6 a^n}
\end{equation}

Defining new co-ordinates $U,V,W,X_2$ using the co-ordinate transformations given earlier ~\ref{coordinatetransformation}, we get the metric of the following form

\begin{displaymath}
ds^2 = \frac{\mu_{1}^{2}}{4} \left( A(V,W,\theta)dV^2 + B(V,W,\theta)dVdU + 2C(V,W,\theta)dVdW + \right.
\end{displaymath}
\begin{equation}
 \left. D(V,W,\theta) dX_{2}^{2} + E(V,W,\theta)dW^2 + F(V,W,\theta)d\Omega_{7}^{2} \right)
\end{equation}

Here the metric coefficients are given by following expressions

\begin{eqnarray}
A(V,W,\theta) =  \frac{f(1-\rho^3)^{-7/3}-1}{V^2 f^{2/3}}  \ \ , \ \  B(V,W,\theta)= \frac{W^2}{f^{2/3}}  \\
C(V,W,\theta) =  \frac{f(1-\rho^3)^{-7/3}-1}{V Wf^{2/3}}  \ \ , \ \  D(V,W,\theta)= \frac{W^2}{f^{2/3}}  \\
E(V,W,\theta) =  \frac{f^{1/3}(1-\rho^3)^{-7/3}}{W^2}  \ \ , \ \  F(V,W,\theta)= \frac{f^{1/3}}{1-\rho^3}  
\end{eqnarray}

we are interested to checking how differentiable are the metric coefficients at the horizon. Integer powers of $V$ don't create any problems and so we concentrate on non-integer powers. 

Knowing the expression for $f$, we can see that the metric is $\mathcal{C}^{1}$ as $V \rightarrow 0$. But we can get a $\mathcal{C}^2$ metric by noting that 
 the lowest non-integer power of $V$ occurs in $dV^2$ term ( $V^{3/2}$) and that can be cancelled by making further co-ordinate transformation $U \rightarrow U -G(V,W,\theta)$ tailored to cancel the term proportional to $V^{3/2}$. In fact, by choosing 
\begin{equation}
G(V,\theta,W) = V^{5/2}L_{1}(W,\theta) + V^{7/2}L_{2}(W,\theta) + \ ...
\end{equation}

we can cancel the half integral powers of $V$ in the coefficient of $dV^2$ term. Explicitly, the co-ordinate transformation is 
\begin{equation}
dVdU \rightarrow dUdV - \frac{\partial G(V,\theta,W)}{\partial V}dV^2 -\frac{\partial G(V,\theta,W)}{\partial W}dVdW -\frac{\partial G(V,\theta,W)}{\partial \theta}dVd\theta
\end{equation}

This co-ordinate transformation makes the metric $C^{2}$ but it generates terms proportional to $V^{5/2}$ in coefficients of $dVdW$ and $dVd\theta$. Notice that since $dV^2$ doesn't contain any non-integer powers of $V$, only terms which contain terms proportional to $V^{5/2}$
are coefficients of $dVdW$ and $dVd\theta$.To obtain a $\mathcal{C}^2$ metric at the horizon, we need to get rid of these terms proportional to $V^{5/2}$. To do that, we make further co-ordinate transformations. First, we write $W= e^{P}$ so that $\frac{dW}{W}= dP$. Then we make following co-ordinate transformations 
\begin{equation}
P \rightarrow P - K_{1}(P,\theta)V^{7/2}    \ \ \ , \ \ \ \theta \rightarrow \theta - K_{2}(P,\theta)V^{7/2}
\end{equation}

Choosing $K_1$ and $K_2$ carefully we can get rid of terms proportional to $V^{5/2}$. Rest of the terms only have $V^{7/2}$ or higher powers of $V$ and hence metric at the horizon $V=0$ is $\mathcal{C}^3$ in these co-ordinates. 

\section{Conclusions and Outlook}

In this work we analyzed in detail the question of degree of smoothness of multiple $M2$-brane metric. By solving for null geodesics in this geometry, we  constructed ``nearly Gaussian null co-ordinates'' and found that in terms of these co-ordinates, metric can be extended across 
the horizon but the extended metric is only $\mathcal{C}^3$ at the horizon. We also found an exact set of co-ordinate transformations which take from multiple $M2$-brane metric in isotropic co-ordinates to co-ordinates(different from Gaussian null co-ordinates) in which metric is $\mathcal{C}^3$ at the horizon. 

Finite differentiabilty of the metric at the horizon means that an observer falling through the horizon can detect the presence of horizon through local measurement \cite{Chrusciel}. Finite differentiabilty means that some derivatives of Riemann tensor will blow up at the horizon and these can in principle be observed by an infalling observer.This singularity is very mild but this is unlike the case of single centered $M2$ branes and black holes, which have analytic horizon. It would be interesting to consider the case of a probe $M2$-brane in this spacetime to see if such kind of singularities have some effects. It's not clear if the fact that interior metric of multiple $M2$-branes is 
not unique has any significance for the world-volume theory. 

One may wonder whether higher derivative corrections have any effect on such singularity. In \cite{Candlish2}, a particular four derivative 
term was considered for the case of multi-centered black holes in five dimensional supergravity and found not to change the degree of smoothness of the metric at the horizon. It's not clear what happens if other higher derivative terms are included. For our M-theory case, higher derivative terms are either not well understood or very difficult to analyze. But the fact that corrections to classical two derivative theory can happen at the scale of the horizon is quite significant in itself. 

The present work also leads to the question of the degree of smoothness of
horizons when the multiple membranes are not confined to a common axis in
the transverse space. One could for example consider multiple membranes
confined to be only on a plane in the transverse space. The exterior
metric would then have only a $SO(6)$ symmetry. The author of 
\cite{Candlish2} argues that the degree of smoothness should decrease. The
more symmetries of the single centered solution that the multi centered
solution breaks the less is the degree of horizon smoothness. Although the
above statement is not a theorem it seems reasonable. We are currently analyzing this issue in the context of multi-black holes and multi-membrane solutions. We plan to report on this issue in a forthcoming work \cite{future}.

\appendix
\section{Series expansions for $r(\lambda, \Theta)$ and $\theta(\lambda, \Theta)$}
In this appendix we give a brief account of the computations leading to \eqref{9} and \eqref{10}. We plug in the expansions \eqref{series} into the equations \eqref{4} and \eqref{5} and solve it order by order in $\lambda$ \cite{Candlish1}. Instead of imposing the boundary conditions \eqref{bc}, we impose the modified boundary conditions :
\be \label{bcn}  r(0, \Theta) = 0, \quad \theta(0, \Theta) = \Theta, \quad \dot{\theta}(0, \Theta) = b. \ee
Doing this ensures that every $c_i$ and $b_i$ co-efficient at whatever order it appears first appears linearly thus giving unique solution. They do occur at higher orders with higher powers but by then they have been determined and become constants in terms of which the solutions to whatever co-efficients that occur  linearly at that order are determined. We do this order by order and then take the limit $b \rightarrow 0$ to obtain the following series expansions. Note : $c_\alpha \equiv \cos \alpha, s_\alpha \equiv \sin \alpha$

\begin{multline}
r(\lambda, \Theta)=\sqrt{2}\mu_ 1^{1/12}\,\lambda^{1/2}+\frac{h_0\,}{3\sqrt{2}\mu_1^{5/12}}\lambda^{7/2}+\frac{16\,h_1}{9\mu_1^{1/3}}\, c_\Theta \,\lambda^ 4 + \frac{4\sqrt{2}\,h_2}{5\,\mu_1^{1/4}}\,(3 + 4 c_{2 \Theta})\,\lambda^{9/2} + \frac{64\,h_3}{33\,\mu_1^{1/6}}\,(9 c_\Theta +5 c_{3\Theta})\lambda^ 5  \\  +\frac{8\sqrt{2}\,h_4}{3\,\mu_1^{1/12}}\,(6+10 c_{2\Theta}+5 c_{4\Theta})\lambda^{11/2} + \frac{64\,h_5}{13}(20 c_\Theta +15 c_{3\Theta}+7 c_{5\Theta})\lambda^6 \nonumber
\end{multline}

\begin{multline}
+\frac{1}{252\sqrt{2}\,\mu_1^{11/12}}(-119\, h_0^2 +38400 \,h_6 \mu _1 + 69120 \,  h_6 \mu _1\,c_{2 \Theta}  + 48384  \,h_6 \mu _1\, c_{4 \Theta}+21504 \, h_6 \mu _1\, c_{6 \Theta})\lambda^{13/2} 
\\
+\frac{64}{135\,\mu_1^{5/6}} (-11\,h_0 h_1\, c_\Theta + 900 \, h_7 \mu _1\,c_{\Theta} + 756 \, h_7 \mu _1 c_{3 \Theta}+504 h_7 \mu _1\,c_{5 \Theta}+216 \,
   h_7 \mu _1 c_{7 \Theta})\,\lambda ^7 
   \\
+\frac{4\sqrt{2}}{14175\mu_1^{3/4}}[-27149\,h_1^2 - 25515 \,h_0h_2+1063125\,h_8 \mu_1 + c_{2 \Theta}(-24476 h_1^2-34020 h_0 h_2+1984500 h_8 \mu _1) 
\\
 +1587600\,h_8 \mu_1 c_{4 \Theta}+1020600\, h_8 \mu_1  c_{6\Theta} 
+ 425250 \,h_8 \mu_1 c_{8 \Theta}]\lambda^{15/2}+ \nonumber
\\
+\frac{32\,}{58905\,\mu_1^{2/3}} [ c_\Theta(-257895 h_1 h_2-97650 h_0 h_3+2910600 h_9 \mu _1) + c_{3 \Theta} (-91377 h_1 h_2-54250 h_0 h_3+2587200 h_9 \mu _1) 
\\
+ 1995840 \, h_9 \mu _1\,c_{5 \Theta}+1247400 \,h_9 \mu _1 c_{7 \,\Theta}+508200\, h_9 \mu _1 c_{9 \Theta}]\lambda^8 \nonumber \\
+ \frac{4\sqrt{2}}{51975\,\mu_1^{7/12}} [ (13582800 \, h_{10} \mu _1 - 1498266 \, h_2^2 - 2034240 \, h_1 h_3 - 635250 \, h_0 h_4) 
+  c_{2 \Theta}(-1984752\, h_2^2-3103360 \,h_1 h_3 
\\ 
-1058750 \,h_0 h_4+25872000 \,h_{10} \mu _1)  +  c_{4 \Theta}\,(-569184 \,h_2^2 - 977600 \, h_1 h_3 - 529375\, h_0 h_4 + 22176000 \, h_{10} \mu _1) 
\\
+ 16632000  \,h_{10}\, \mu _1 c_{6 \Theta} + 10164000 \, h_{10} \mu _1 c_{8 \, \Theta} + 4065600 \, h_{10} \mu _1\,c_{10\,\Theta}]\lambda^{17/2} 
\\
+\frac{64}{513513 \, \mu _1^\frac12} [ c_\Theta (-14774760 \,h_2 h_3-8500349\, h_1 h_4-2411640\, h_0 h_5+42378336\, h_{11} \mu _1) + c_{3 \Theta} (-8162154\, h_2 h_3
\\
-5645640\, h_1 h_4 - 1808730\,  h_0 h_5 + 38918880\, h_{11} \mu _1)  +  c_{5 \Theta}(-2104830\, h_2 h_3-1661803 \,h_1 h_4-844074\, h_0 h_5
\\
+32432400\, h_{11} \mu _1) + 23783760  \,h_{11} \mu _1 c_{7 \Theta} + 14270256 \, h_{11} \mu _1 c_{9
\Theta} + 5621616\,  h_{11} \mu _1 c_{11 \Theta}]  \lambda^9
\\
+\frac{1}{249729480 \sqrt{2} \,\mu _1^{17/12}} [(306997691 \,h_0^3 - 115430515200 \,h_6 \mu _1 h_0 + 1670723518464\, h_{12} \mu _1^2 - 569874898944\, h_3^2 \mu _1 
\\
- 895030456320\, h_2 h_4 \mu _1 - 446778630144\,    h_1 h_5 \mu _1) + c_{2 \Theta} (-881051738112  \,h_3^2 \mu _1 + 3222109642752\, h_{12} \mu _1^2  
\\
- 1476800252928\,  h_2 h_4 \mu _1 - 776493527040\, h_1 h_5 \mu _1 - 207774927360\, h_0 h_6 \mu _1 ) + c_{4 \Theta} (-417291264000 \, h_3^2 \mu _1 
\\
+ 2876883609600 \,h_{12} \mu _1^2 - 738400126464\, h_2 h_4 \mu _1 - 474275450880\, h_1 h_5 \mu _1 - 145442449152 \,h_0 h_6 \mu _1) 
\\
+ c_{6 \Theta} (-97068441600\,h_3^2  \mu _1  + 2344127385600\, h_{12} \mu _1^2 - 179006091264\, h_2 h_4 \mu _1 - 133822402560\, h_1 h_5 \mu _1 
\\
- 64641088512\, h_0 h_6 \mu _1) +  1687771717632 \, h_{12} \mu _1^2 \,c_{8 \Theta}  +  997319651328\, h_{12} \mu _1^2 \,c_{10 \Theta} + 387846531072\, h_{12} \mu _1^2 \, c_{12 \Theta}] \lambda^\frac92 
\\
+\frac{32\,}{14189175 \mu _1^{4/3}}[c_\Theta\,(9214205 \,h_1 h_0^2 - 567567000 \,h_7 \mu _1 h_0 + 7264857600\,  h_{13} \mu _1^2 - 7311447000\,  h_3 h_4 \mu _1 
\\
- 5050854900\, h_2 h_5 \mu _1 - 2327222040\, h_1 h_6
   \mu _1) + c_{3 \Theta} (6810804000 \,h_{13} \mu _1^2 - 4918914000\,  h_3 h_4 \mu _1 - 3736223820 \, h_2 h_5 \mu _1
\\  
- 1855812816\,  h_1 h_6 \mu _1 - 476756280\,  h_0 h_7 \mu _1) + c_{5 \Theta}\,(5945940000\, h_{13} \mu _1^2 - 2134875600\, h_3 h_4 \mu _1 - 1756305180\, h_2 h_5 \mu _1 
\\
- 1072119048 \,h_1 h_6 \mu _1 - 317837520\, h_0 h_7 \mu _1) + c_{7 \Theta}(4756752000\, h_{13} \mu _1^2 - 467181000\, h_3 h_4 \mu _1 - 409786020\, h_2 h_5 \mu _1 
\\
- 294390096\, h_1 h_6 \mu _1 - 136216080\, h_0 h_7 \mu _1) + 3372969600 \,h_{13} \mu _1^2 \, c_{9 \Theta} + 1967565600 \,h_{13} \mu _1^2\, c_{11 \Theta} 
\\
+ 756756000 \, h_{13} \mu _1^2\,c_{13 \Theta}]\lambda^{10} +\mathcal{O}(\lambda^{11})
\end{multline}

\begin{multline}
\theta(\lambda, \Theta) = \Theta -\frac{24\sqrt{2}h_1\,}{35\mu_1^{5/12}} s_\Theta \,\lambda^{7/2}- \frac{4 h_2\, \,}{\mu_1^{1/3}} s_{2 \Theta}\,\lambda^4 -\frac{32\,\sqrt{2}\,h_3\,}{21\mu_1^{1/4}}\,(3 s_\Theta +5 s_{3\Theta})\,\lambda^{9/2}\\ -\frac{24\,h_4}{\mu_1^{1/6}}\,(s_{2 \Theta} + s_{4 \Theta})\,\lambda^5 
-\frac{160\,\sqrt{2}\,h_5}{33\,\mu_1^{1/12}}\,(4 s_\Theta + 9 s_{ 3\Theta} +7 s_{5\Theta})\,\lambda^{11/2} -\frac{32\,h_6\,\,}{5}\,(15\,s_{ 2 \Theta} + 21s_{4 \Theta} + 14 s_{6 \Theta})\,\lambda^6 \\ +\frac{4 \sqrt{2}  }{715 \mu _1^{11/12}}\,(241\,  h_0 h_1 s_\Theta-12000\, h_7 \mu _1 s_\Theta  -30240\,h_7 \mu _1 s_{3 \Theta}  -33600\,h_7 \mu _1 s_{5
   \Theta}  -20160 \,h_7 \mu _1  s_{7 \Theta})\,\lambda^\frac{13}{2} \\+ \frac{16\, \,}{735 \,\mu _1^{5/6}}\,(344 \, h_1^2 s_{2 \Theta} +385 \, h_0 h_2\,s_{2 \Theta} - 14700  \,h_8 \mu _1\, s_{2 \Theta}-23520\, h_8 \mu _1\,s_{4 \Theta} - 22680 \, h_8 \mu _1\,s_{6 \Theta} - 12600 \, h_8 \mu _1\,s_{8 \Theta})\lambda^7 \\ + \frac{16 \sqrt{2}}{20475 \,\mu _1^{3/4}} ( 37825\,h_1 h_2\, s_\Theta  + 48247  \,h_1 h_2\, s_{3 \Theta} + 12825 \,  h_0 h_3\, s_\Theta + 21375\,  h_0 h_3\, s_{3 \Theta} - 264600 \, 
   h_9 \mu _1\, s_\Theta \\ -705600\,   h_9 \mu _1\, s_{3 \Theta}-907200\,  h_9 \mu _1\,s_{5 \Theta} -793800\,  h_9 \mu _1\, s_{7 \Theta} - 415800 \,  h_9 \mu _1\, s_{9 \Theta} ) \lambda^{\frac{15}{2}} + \mathcal{O}(\lambda^8) \nonumber
\end{multline}

\section{Components of the $\mathcal{C}^3$ metric}
Here we gather the formulae for various terms defined in \eqref{c3metric}
\begin{eqnarray}\label{b1}
H^{-\frac{2}{3}}(\lambda, \Theta) &=& \frac{4  }{ \mu _1^{1/3}} \lambda^2 -\frac{56\, h_0}{3 \mu _1^{5/6}}\lambda ^5 -\frac{1024 \sqrt{2}  \cos \Theta \, h_1}{9 \mu _1^{3/4}}\lambda ^{11/2} + \mathcal{O}(\lambda^{6}) \nonumber \\
2T(\lambda, \Theta)H^{-\frac{2}{3}}(\lambda, \Theta) &=& -2\lambda +\frac{14 \lambda ^4 h_0}{\sqrt{\mu_1}}+\frac{3584 \sqrt{2} \lambda ^{9/2} \cos \Theta  h_1}{45 \mu_1^{5/12}} + \mathcal{O}(\lambda^{5}) \nonumber \\
T^2(\lambda, \Theta)H^{-\frac{2}{3}}(\lambda, \Theta) &=& \frac{\mu_1^{1/3}}{4}-\frac{7 \lambda ^3 h_0}{3 \mu_1^{1/6}}-\frac{64 \sqrt{2} \lambda ^{7/2} \cos \Theta h_1}{5
\mu_1^{1/12}} + \mathcal{O}(\lambda^{4}) \nonumber \\
2\partial_{\Theta}T(\lambda, \Theta)H^{-\frac{2}{3}}(\lambda,\Theta) &=& -\frac{1024 \sqrt{2} \lambda ^{9/2} h_1 \sin \Theta}{45 \mu_1^{5/12}}-\frac{768 \lambda ^5 h_2 \sin(2 \Theta)}{5 \mu_1^{1/3}}+ \mathcal{O}(\lambda^{11/2}) \nonumber \\
H^{\frac{1}{3}}(\lambda,\Theta)r^2(\lambda,\Theta) &=& \mu_1^\frac13+\frac{8 h_0 }{3 \mu _1^{1/6}} \lambda ^3  +\frac{16 \sqrt{2}  \cos \Theta  h_1 }{ \mu _1^{1/12}}\lambda ^{7/2} +\mathcal{O}(\lambda^4) \nonumber \\
H^{\frac{1}{3}}(\lambda,\Theta)r^2(\lambda,\Theta)\sin ^2(\theta(\lambda,\Theta)) &=& \sin ^2 \Theta \, \left(\mu _1^\frac13+\frac{8  h_0}{3   \mu _1^{1/6}} \lambda^3 +\frac{512 \sqrt{2}  \cos \Theta   h_1 }{35 \mu_1^{1/12}}\, \lambda^\frac72 +\mathcal{O}(\lambda^4)\right) \nonumber \\
\end{eqnarray}

\begin{eqnarray} \label{b2}
z_1(v,w,y_2) & \equiv &  -\left(\partial_v f \right)^2+\left(\partial_v g \right)^2+\left(\partial_v h \right)^2   \nonumber \\
z_2(v,w,y_2) & \equiv &  -\left(\partial_w f \right)^2+\left(\partial_w g \right)^2+\left(\partial_w h \right)^2  \nonumber \\
z_3(v,w,y_2) & \equiv &  -\left(\partial_{y_2} f \right)^2+\left(\partial_{y_2} g \right)^2+\left(\partial_{y_2} h \right)^2  \nonumber \\
q_1(v,w,y_2) & \equiv &  \partial_w f - \partial_v g \nonumber \\
q_3(v,w,y_2) & \equiv &  \partial_{y_2} f - \partial_v h  \nonumber \\
q_5(v,w,y_2) & \equiv & - \left(\partial_w h + \partial_{y_2} g\right) \nonumber \\
q_2(v,w,y_2) & \equiv &  -\partial_v f \, \partial_w f+ \partial_v g \, \partial_w g + \partial_v h \, \partial_w h   \nonumber \\
q_4(v,w,y_2) & \equiv &  -\partial_v f \, \partial_{y_2} f+ \partial_v g \, \partial_{y_2} g + \partial_v h \, \partial_{y_2} h  \nonumber \\
q_6(v,w,y_2) & \equiv &  -\partial_{y_2} f \, \partial_w f+ \partial_{y_2} g \, \partial_w g + \partial_{y_2} h \, \partial_w h  
\end{eqnarray}

\section{Components of $A_{[3]}$}
In this appendix, the components of the 3-form potential \eqref{mpot} are explicitly given in \eqref{c1},  \eqref{c21} and \eqref{c22}.. The various terms defined in the 4-form field strength \eqref{mfield} are in \eqref{c21}, \eqref{c22}  and \eqref{c3}. 

\begin{eqnarray} \label{c3}
u_1(v,w,y_2) & \equiv & \partial_{y_2}h + \partial_v f + \partial_w g \nonumber \\
u_3(v,w,y_2) & \equiv & h \, \partial_v f - f \, \partial_v h + h \, \partial_w g -g \, \partial_w h \nonumber \\
u_5(v,w,y_2) & \equiv & f \, \partial_v g - g \, \partial_v f + h \, \partial_{y_2} g -g \, \partial_{y_2} h \nonumber \\
u_7(v,w,y_2) & \equiv & f \, \partial_{y_2} h - h \, \partial_{y_2} f + f \, \partial_w g -g \, \partial_w f \nonumber \\
u_2(v,w,y_2) & \equiv & \left(\partial_v f \, \partial_{y_2}h - \partial_{y_2}f \, \partial_v h \right) + \left(\partial_w g \, \partial_{y_2}h - \partial_{y_2}g \, \partial_w h \right)  + \left(\partial_v f \, \partial_{w}g - \partial_{w}f \, \partial_v g \right) \nonumber \\
u_4(v,w,y_2) & \equiv & f \, \left(\partial_v h \, \partial_w g - \partial_w h \, \partial_v g \right) + g \, \left(\partial_v f \, \partial_w h - \partial_w f \, \partial_v h \right) + h \, \left(\partial_v g \, \partial_w f - \partial_w g \, \partial_v f \right) \nonumber \\
u_6(v,w,y_2) & \equiv & f \, \left(\partial_v h \, \partial_{y_2} g - \partial_{y_2} h \, \partial_v g \right) + g \, \left(\partial_v f \, \partial_{y_2} h - \partial_{y_2} f \, \partial_v h \right)  + h \, \left(\partial_v g \, \partial_{y_2} f - \partial_{y_2} g \, \partial_v f \right) \nonumber \\
u_8(v,w,y_2) & \equiv & f \, \left(\partial_w h \, \partial_{y_2} g - \partial_{y_2} h \, \partial_w g \right) + g \, \left(\partial_w f \, \partial_{y_2} h - \partial_{y_2} f \, \partial_w h \right) + h \, \left(\partial_w g \, \partial_{y_2} f - \partial_{y_2} g \, \partial_w f \right) \nonumber \\
\end{eqnarray}

\begin{eqnarray}\label{c1}
A_{[3]} &=& H^{-1}dt \wedge dx_1 \wedge dx_2 \nonumber \\
&=& H^{-1} \left( dv-fdT-Tdf \right) \wedge \left( dw-gdT -Tdg \right) \wedge \left( dy_2 -hdT -Tdh \right) \nonumber \\
&=& H^{-1}(\lambda,\Theta) \, dv \wedge dw \wedge dy_2 
 - H^{-1}(\lambda,\Theta) \, \partial_{\lambda}T(\lambda,\Theta) \, \Big[ h \,  d\lambda \wedge dv\wedge dw +f \,  d\lambda \wedge dw \wedge dy_2 + 
\nonumber \\ && g \,  d\lambda \wedge dv \wedge dy_2 \Big] - \mathcal{O}(\lambda^{11/2}) \, \Big[ h \,  d\Theta \wedge dv\wedge dw +f \,  d\Theta \wedge dw \wedge dy_2 + g \,  d\Theta \wedge dv \wedge dy_2 \Big]
\nonumber \\ &&-H^{-1}(\lambda,\Theta)\, T(\lambda,\Theta) \, \Big[ dv \wedge dw \wedge dh + df \wedge dw \wedge dy_2 + dv \wedge dg \wedge dy_2 \Big]
\nonumber \\ &&+H^{-1}(\lambda,\Theta) \, T(\lambda,\Theta) \, \partial_{\lambda}T(\lambda,\Theta) \, \Big[ f \, \left( d\lambda \wedge dw \wedge dh + d\lambda \wedge dg \wedge dy_2 \right) 
\nonumber \\ &&+g \, \left( d\lambda \wedge dy_2 \wedge df + d\lambda \wedge dh \wedge dv \right)+h \, \left( d\lambda \wedge dv \wedge dg + d\lambda \wedge df \wedge dw \right) \Big]
\nonumber \\ &&+ \mathcal{O}(\lambda^{9/2}) \, \Big[ f \, \left( d\Theta \wedge dw \wedge dh + d\Theta \wedge dg \wedge dy_2\right)
\nonumber \\ &&+g \, \left( d\Theta \wedge dy_2 \wedge df + d\lambda \wedge dh \wedge dv \right)+h \, \left( d\Theta \wedge dv \wedge dg + d\lambda \wedge df \wedge dw \right) \Big]
\nonumber \\ &&+ H^{-1}(\lambda,\Theta) \, T^2(\lambda,\Theta) \, \Big[df \wedge dw \wedge dh + dv \wedge dg \wedge dh + df \wedge dg \wedge dy_2\Big]
\nonumber \\ &&+ H^{-1}(\lambda,\Theta) \, T^2(\lambda,\Theta) \, \partial_{\lambda}T(\lambda,\Theta) \, \Big[f \, d\lambda \wedge dh \wedge dg + g \, d\lambda \wedge df \wedge dh + h \, d\lambda \wedge dg \wedge df \Big]
\nonumber \\ &&+ H^{-1}(\lambda,\Theta) \, T^2(\lambda,\Theta) \, \partial_{\Theta}T(\lambda,\Theta) \, \Big[f \, d\Theta \wedge dh \wedge dg + g \, d\Theta \wedge df \wedge dh + h \, d\Theta \wedge dg \wedge df \Big] \nonumber \\
\end{eqnarray}

Various combinations which occur in \eqref{mfield} having following series expansions in $\lambda$.
\begin{eqnarray}\label{c21}
\partial_{\lambda}H^{-1} &=& \frac{24}{\sqrt{ \mu_1}} \lambda ^2-\frac{336  h_0}{ \mu_1 }\lambda ^5-\frac{6656 \sqrt{2}  \cos \Theta h_1}{3  \mu_1^{11/12}}\lambda ^{11/2}+\mathcal{O}(\lambda^6) \nonumber \\
T \, \partial_{\lambda}H^{-1} &=& -\frac{6 }{ \mu_1 ^{1/6}}\lambda+\frac{98  h_0}{ \mu_1 ^{2/3}}\lambda^4+\frac{9344 \sqrt{2}  \cos \Theta h_1}{15 \mu_1 ^{7/12}}\lambda ^{9/2} +\mathcal{O}(\lambda^5) \nonumber \\
T^2 \, \partial_{\lambda}H^{-1} &=& \frac{3 \mu_1^{1/6}}{2}-\frac{28  h_0}{ \mu_1^{1/3}}\lambda ^3-\frac{864 \sqrt{2}  \cos \Theta h_1}{5 \mu_1 ^{1/4}} \lambda ^{7/2}+ \mathcal{O}(\lambda^4) \nonumber \\
\partial_{\Theta}H^{-1} &=& \frac{1024 \sqrt{2}  h_1 \sin \Theta }{3  \mu_1 ^{11/12}} \lambda ^{13/2} + \mathcal{O}(\lambda^7) \nonumber \\
T \, \partial_{\Theta}H^{-1} &=& -\frac{256 \sqrt{2}  h_1 \sin \Theta }{3 \mu_1^{7/12}} \lambda^{11/2} + \mathcal{O}(\lambda^6) \nonumber \\
T^2 \, \partial_{\Theta}H^{-1} &=& \frac{64 \sqrt{2} h_1 \sin \Theta }{3  \mu_1 ^{1/4}} \lambda ^{9/2}  + \mathcal{O}(\lambda^5)  
\end{eqnarray}

\begin{eqnarray} \label{c22}
\partial_{\lambda}H^{-1} \, \partial_{\Theta}T - \partial_{\lambda}T \, \partial_{\Theta}H^{-1} &=& \frac{768 \sqrt{2}  h_1 \sin \Theta }{5 \mu_1 ^{7/12}} \lambda ^{9/2}+ \mathcal{O}(\lambda^5) \nonumber \\
T \, \left( \partial_{\lambda}H^{-1} \, \partial_{\Theta}T - \partial_{\lambda}T \, \partial_{\Theta}H^{-1}\right) &=& -\frac{192 \sqrt{2} h_1 \sin \Theta }{5 \mu_1 ^{1/4}} \lambda ^{7/2}  + \mathcal{O}(\lambda^4) \nonumber \\
T^2 \, \left( \partial_{\lambda}H^{-1} \, \partial_{\Theta}T - \partial_{\lambda}T \, \partial_{\Theta}H^{-1}\right) &=& \frac{48}{5} \sqrt{2}  \mu_1^{1/12} h_1 \sin \Theta \,\lambda ^{5/2}  + \mathcal{O}(\lambda^3) 
\end{eqnarray}

\section{Equation of motion}

In this appendix we gather all the formulae that goes into the computations of section 4.5. We work with a specific choice of Gaussian null-like co-ordinates:
\begin{eqnarray}\label{d1}
t &=& v-f(w,y_2) \, T(\lambda , \Theta) \nonumber \\
x_1 &=& v-g(w,y_2) \, T(\lambda , \Theta) \nonumber \\
x_2 &=& y_2 \, T(\lambda , \Theta)
\end{eqnarray}
where
\begin{eqnarray}
f(w,y_2) &\equiv & \frac{1}{2}\left( w + \frac{1}{w} + \frac{y_2^2}{w} \right) \nonumber \\
g(w,y_2) &\equiv & \frac{1}{2}\left( -w + \frac{1}{w} + \frac{y_2^2}{w} \right). 
\end{eqnarray}
In this specific Gaussian null-like co-ordinate system the 4-form field strength is 
\ben
F_{[4]}=(\partial_{\lambda}H^{-1}d\lambda + \partial_{\Theta}H^{-1}d\Theta ) \wedge \Bigg[(f-g)Tdv \wedge dT \wedge dy_2 -y_2 T dv \wedge dg \wedge dT - T^2dv \wedge dg \wedge dy_2 \nonumber \\ -T^2df \wedge dv \wedge dy_2 + fT^2 dT \wedge dg \wedge dy_2 -gT^2 dT \wedge df \wedge dy_2 -y_2T^2 df \wedge dg \wedge dT - y_2 T df \wedge dv \wedge dT \Bigg], \nonumber
\een

\ben \label{d5}
F_{[4]}=\left( \frac{48 \sqrt{2} \mu_1^{1/12} h_1 \sin \Theta }{5 w}  \lambda ^{5/2}+\frac{72 \mu_1^{1/6} h_2 \sin (2 \Theta)}{w}  \lambda ^3+\mathcal{O}(\lambda^{7/2}) \right) d\lambda \wedge d\Theta \wedge dw \wedge dy_2 \nonumber \\
+\left( \frac{3 \mu_1^{1/6}}{2}-\frac{28 h_0 }{\mu_1^{1/3}} \lambda ^3-\frac{864 \sqrt{2} \cos \Theta  h_1 }{5 \mu_1^{1/4}} \lambda ^{7/2}+\mathcal{O}(\lambda^4) \right) d\lambda \wedge dv \wedge dw \wedge dy_2 \nonumber \\
+\left( \frac{64 \sqrt{2} h_1 \sin \Theta }{3  \mu_1^{1/4}} \lambda ^{9/2}+\mathcal{O}(\lambda^5) \right) d\Theta \wedge dv \wedge dw \wedge dy_2 \nonumber \\
+\left(-\frac{192 \sqrt{2} y_2 h_1 \sin \Theta }{5 \mu_1^{1/4}} \lambda ^{7/2}-\frac{288 y_2 h_2 \sin(2\Theta) }{\text{$\mu $1}^{1/6}} \lambda ^4 +\mathcal{O}(\lambda^{7/2}) \right) d\lambda \wedge d\Theta \wedge dv \wedge dw \nonumber \\
\left( \frac{768 \sqrt{2} w h_1 \sin \Theta}{5 \mu_1^{7/12}}\lambda ^{9/2} +\mathcal{O}(\lambda^5)\right)d\lambda \wedge d\Theta \wedge dv \wedge dy_2. \nn
\een
The non-zero components of the Ricci tensor (upto and \emph{including} $\mathcal{O}(\lambda^{3})$ ) are listed here:

\begin{eqnarray}\label{ricci}
R_{vw} &=& \frac{12  }{\mu_1^{1/3}} \lambda +\mathcal{O}(\lambda^{7/2}) \nonumber \\
R_{ww} &=& -\frac{3 \left(1+ y_2^2\right)}{w^2}+\frac{84 \left(1+y_2^2\right) h_0}{w^2 \sqrt{\mu_1}}\lambda ^3 +\mathcal{O}(\lambda^{7/2}) \nonumber \\
R_{wy_2} &=& \frac{3 y_2}{w}-\frac{84 y_2  h_0}{w \sqrt{\mu_1}} \lambda ^3 +\mathcal{O}(\lambda^{7/2})\nonumber \\
R_{y_2y_2} &=& -3+\frac{84  h_0}{\sqrt{\mu_1}} \lambda ^3 +\mathcal{O}(\lambda^{7/2}) \nonumber \\
R_{v\lambda } &=& -\frac{12 w}{\mu_1^{1/3}}+\frac{224 w  h_0}{\mu_1^{5/6}}\lambda ^3 +\mathcal{O}(\lambda^{7/2})\nonumber \\
R_{\lambda \Theta}  &=& -\frac{576 \sqrt{2} h_1 \sin \Theta }{5 \mu_1^{5/12}} \lambda ^{5/2}  -\frac{1728 h_2 \cos \Theta \sin \Theta }{\mu_1^{1/3}} \lambda ^3 +\mathcal{O}(\lambda^{7/2})\nonumber \\
R_{\Theta \Theta}  &=& 6-\frac{96  h_0}{\sqrt{\mu_1}} \lambda ^3  +\mathcal{O}(\lambda^{7/2}) \nonumber \\
R_{a b} &=& 6 \left(1-\frac{16  h_0}{\sqrt{\mu_1}} \lambda ^3 +\mathcal{O}(\lambda^{7/2}) \right) \sin^2\, \Theta \, g_{ab},
\end{eqnarray}

where in the last equation $R_{ab}$ and $g_{ab}$ are the components of the Ricci tensor and the metric tensor along the six sphere. Other components may 
be non-zero but they all start after $\mathcal{O}(\lambda^{3})$.


\begin{thebibliography}{99}


\bibitem{Candlish1}
  G.~N.~Candlish and H.~S.~Reall,
  ``On the smoothness of static multi-black hole solutions of
  higher-dimensional Einstein-Maxwell theory,''
  Class.\ Quant.\ Grav.\  {\bf 24}, 6025 (2007)
  [arXiv:0707.4420 [gr-qc]].

\bibitem{Candlish2}
  G.~N.~Candlish,
  ``On the smoothness of the multi-BMPV black hole spacetime,''
  Class.\ Quant.\ Grav.\  {\bf 27}, 065005 (2010)
  [arXiv:0904.3885 [hep-th]].

\bibitem{Gibbons}
  G.~W.~Gibbons, G.~T.~Horowitz and P.~K.~Townsend,
  ``Higher Dimensional Resolution Of Dilatonic Black Hole Singularities,''
  Class.\ Quant.\ Grav.\  {\bf 12}, 297 (1995)
  [arXiv:hep-th/9410073].

\bibitem{Welch}
  D.~L.~Welch,
  ``On the smoothness of the horizons of multi - black hole solutions,''
  Phys.\ Rev.\  D {\bf 52}, 985 (1995)
  [arXiv:hep-th/9502146].

\bibitem{Hartle}
  J.~B.~Hartle and S.~W.~Hawking,
  ``Solutions of the Einstein-Maxwell equations with many black holes,''
  Commun.\ Math.\ Phys.\  {\bf 26}, 87 (1972).
\bibitem{Duff} 
  M.~J.~Duff and K.~S.~Stelle,
  ``Multimembrane solutions of D = 11 supergravity,''
  Phys.\ Lett.\ B {\bf 253}, 113 (1991).

\bibitem{Stelle}
  K.~S.~Stelle,
  ``BPS branes in supergravity,''
  arXiv:hep-th/9803116.

\bibitem{Perry} 
  C.~Codirla and M.~J.~Perry,
  ``Compactification of supermembranes,''
  Nucl.\ Phys.\ B {\bf 561}, 43 (1999)
  [hep-th/9809043].

\bibitem{Myers} 
  R.~C.~Myers,
  ``Higher Dimensional Black Holes In Compactified Space-times,''
  Phys.\ Rev.\ D {\bf 35}, 455 (1987).

\bibitem{Chrusciel} 
  P.~T.~Chrusciel and D.~B.~Singleton,
  ``Nonsmoothness of event horizons of Robinson-Trautman black holes,''
  Commun.\ Math.\ Phys.\  {\bf 147}, 137 (1992).

\bibitem{future}
 Chethan N. Gowdigere, Siddharth Satpathy, Yogesh K. Srivastava, work in
progress.


\end{thebibliography}
\end{document}